**Towards $N$ mode parametric electromechanical resonances**

Authors: Adarsh Ganesan[1], Cuong Do[1], Ashwin Seshia[1]

[1] Nanoscience Centre, University of Cambridge, Cambridge, UK, CB3 0FF

**The ubiquity of parametric resonance is continually evident in the repeated experimental observations of this phenomenon in multiple physical systems [1,3-19]. The elementary case of 2 mode parametric resonance of order 1 involves the excitation of a spectral tone of a parametrically driven mode at a sub-harmonic frequency of the higher directly driven mode. Historically, such examples of parametric resonance have been predominantly researched in a system of micro- and nanoelectromechanical resonators. Here, in this paper, we break this convention by showcasing a collection of experimental signatures in support of the concept of '$N$ mode parametric resonance' using a number of elementary microelectromechanical devices. Specifically, we present observations of 2, 3, $(2+3)$ and $(3+3)$ mode parametric resonances demonstrating co-existence of different regimes within the same device. In addition, we also present observations of intrinsic 'Four-Wave Mixing' of parametric excitations. This paper presents contributions towards the existence proof for such multimode parametric resonances which can also be exploited for engineering benefit within the field of 'micro and nanoelectromechanical resonators'. The experimental results further point towards the possibility of the ultimate observation of $N$ mode parametric resonance in such physical system.**

Michael Faraday's original observations of parametric resonance [1] have led to significant follow-on research and engineering application, establishing this discovery as one of the most remarkable and well received nonlinear mechanisms. This phenomenon relies on the periodic modulation of the system parameters of a second-order harmonic oscillator, with the onset of the fundamental mode parametric resonance occurring when the system is driven at twice the natural frequency as described by the Mathieu framework [2]. The phenomenon of parametric resonance has also been demonstrated in a number of disparate physical systems including elementary particles [3-4], astrophysics [5], fluid mechanics [1,6-8], magnetism [9], electronics [10], optical lasers [11-12], mechanics [13] and biophysics [14-15]. More recently, with the advancements in micro- and nanofabrication technology, parametric resonance has also been successfully observed in micro and nanoelectromechanical systems [16-19] and researched for applications to sensors, low-noise oscillators and vibration energy harvesting. The enhanced amplitude of oscillation relative to ordinary resonance is one of the intriguing qualities of parametric resonance and amplitude stabilization is limited by system nonlinearities which can potentially be exquisitely engineered and controlled.

Conventionally, parametric resonance is simply associated with sub-harmonic excitation of a specific vibration mode owing to the popularity of such observations [1,3-19]. However, it is theoretically conceivable for a non-linearly coupled dynamical system to exhibit higher order $N$ mode parametric resonance wherein $N$ separate vibration modes (with individual resonances at $\approx f_i$) are excited by a drive tone $f_d$ such that $f_d = \sum_{i=1}^{N} \xi_i f_i ; \xi_i \in \mathbb{Z}$. Over and above that, the coexistence of multiple $N$ mode parametric resonances is also theoretically feasible such that many possible permutations of such $N$ mode parametric resonances exist within a single device, particularly at high drive frequencies, high mode number and high modal density. Such a phenomenon will be exhibited in non-linear crystal lattice for instance, where the restoring forces (and as a consquence the non-linear interactions between modes) are approximated by higher order polynomial terms. Despite such a theoretical perception of multimode dynamics, there are only a few experimental systems which exhibit $N$ mode parametric resonance [20-22]. In this paper, we demonstrate a system of microelectromechanical resonators that experimentally validate such a hypothesis for low values of $N$, pointing towards the ultimate possibility of such $N$-mode parametric resonances.

Let us consider an $N$ mode system with each mode being represented by a harmonic oscillator containing the nonlinear terms. The motion of $i^{th}$ oscillator can then be described as

$$\ddot{Q}_i + \omega_i^2 Q_i + 2\zeta_i \omega_i \dot{Q}_i + \underbrace{\sum_{\substack{n=1 \\ n \neq i}}^{N} \mu_{in} Q_i Q_n}_{2 \text{ mode}} + \underbrace{\sum_{\substack{n=1 \\ n \neq i}}^{N} \sum_{\substack{m=1 \\ m \neq i}}^{N} \mu_{nm} Q_n Q_m}_{3 \text{ mode}} + \cdots + \underbrace{\chi_i \prod_{\substack{n=1 \\ n \neq i}}^{N} Q_n}_{N \text{ mode}} + P_i \cos \Omega t = 0; \quad (1)$$

$$i = 1, 2, \ldots, N$$

where $\frac{\omega_i}{2\pi}$ and $\zeta_i$ are the natural frequency and damping coefficient respectively, $\mu_{nm}$ and $\chi_i$ corresponds to the $2^{nd}$ order (or quadratic) and $N^{th}$ order nonlinear coupling coefficients respectively. Such coefficients pave the way for parametric excitations in the system of coupled harmonic oscillators. Specifically, the terms $\mu_{in} Q_i Q_n$, $\mu_{nm} Q_n Q_m$ and $\chi_i \prod_{\substack{n=1 \\ n \neq i}}^{N} Q_n$ in the eq. (1) are responsible for 2 mode, 3 mode and $N$ mode parametric resonances respectively.

The experimental validation was conducted on three micromechanical resonator designs (Figure 1), fabricated in a thin-film piezoelectric-on-SOI microfabrication process [23]. In brief, this includes (i) a free-free beam microstructure of dimensions $1100 \times 350 \times 11\ \mu m^3$ (Figure 1A) (Device 1), (ii) two free-free beam microstructures of dimensions $1100 \times 350 \times 11\ \mu m^3$ coupled by a mechanical coupler of dimensions $20 \times 2 \times 11\ \mu m^3$ (Figure 1B) (Device 2), and three free-free beam microstructures of dimensions $1100 \times 350 \times 11\ \mu m^3$ coupled by mechanical couplers of dimensions $20 \times 2 \times 11\ \mu m^3$ (Figure 1C) (Device 3). The higher transduction efficiencies associated with the piezoelectrically operated micromechanical devices provide access to nonlinear excitations of vibratory modes in these device structures at nominal drive power levels. The electrical drive signals for the device are derived from a function generator and the mechanical motion of the devices is recorded using the Laser Doppler Vibrometry (LDV). All experiments were conducted under ambient room temperature and atmospheric pressure conditions. Device 1 (Figure 1A) was used to demonstrate 2, $(2+2)$ and 3 mode parametric resonances, Device 2 (Figure 1B) was used to demonstrate $(3+3)$ parametric resonance, and Device 3 (Figure 1C) has been used to demonstrate $(2+3)$ parametric resonance. The experimental observations of parametric resonances in these systems are further elaborated below.

Case 1: 2 & $(2+2)$ mode parametric resonances – Let us consider the case where the drive frequency $\frac{\Omega}{2\pi} \cong \frac{\omega_1}{2\pi}$. Based on linear resonance, mode 1 gets directly excited. At elevated drive levels of $P_1$, additionally, the mode $n \in \{2,3,\dots,N\}$ may also get parametric excited via the coupling coefficient $\mu_{1n}$ if the frequency matching condition $\frac{\Omega}{2\pi} \cong \frac{\omega_n}{\pi}$ is satisfied.. The activation threshold for such parametric resonance is determined by the damping coefficient $\zeta_n$ and the magnitude of the coupling coefficient. In situations where multiple modes satisfying the modal frequency ratio $\frac{\omega_1}{\omega_n} \cong 2$ exist, it is certainly possible to have simultaneous excitation of such modes satisfying the frequency matching conditions $\frac{\Omega}{2\pi} \cong \frac{\omega_n}{\pi}$; $n \in \{2,3,\dots,N\}$. The experimental observations in light of this conception are presented in the figure 2A. Here, mode 1 is directly driven at $\frac{\Omega}{2\pi} = 3.874\ MHz \cong \frac{\omega_1}{2\pi}$ couples with the modes of natural frequencies $\cong \frac{\Omega}{4\pi}$ (Figure 2C). For subtle detuning of the drive frequency, $\frac{\Omega}{2\pi} = 3.862\ MHz$ and $\frac{\Omega}{2\pi} = 3.876\ MHz$, , the frequency matching condition switches from $\frac{\Omega}{2\pi} \cong \frac{\omega_2}{\pi}$ to $\frac{\Omega}{2\pi} \cong \frac{\omega_3}{\pi}$ resulting in the preferential parametric excitation of one specific mode over the other (Figures 2B and 2D).

Case 2: 3 mode parametric resonance – Again, we consider the case where the drive frequency $\frac{\Omega}{2\pi} \cong \frac{\omega_1}{2\pi}$. Based on linear resonance, mode $i = 1$ gets directly excited. At elevated drive levels of $P_1$, additionally, modes 2 and 3 may also get excited via 3 mode parametric resonance through the coefficients $\mu_{12}, \mu_{23}$ & $\mu_{13}$ if the frequency matching condition $\frac{\Omega}{2\pi} \cong \frac{\omega_2+\omega_3}{2\pi}$ is satisfied [22, 24]. The experimental data shown in the figure 2E supports this fact. Here, the tones of frequencies $5.387\ MHz \cong \frac{\omega_2}{2\pi}$ and $6.944\ MHz \cong \frac{\omega_3}{2\pi}$ are parametrically excited (Figure 2E) when the mode of frequency $\frac{\omega_1}{2\pi}$ is driven at $\left(\frac{\Omega}{2\pi} = 12.331\ MHz\right) \cong \frac{\omega_1}{2\pi} \cong \frac{\omega_2+\omega_3}{2\pi}$ and at a higher drive level of $18\ dBm$. These tones correspond to two different eigenmodes with natural frequencies $\frac{\omega_2}{2\pi}$ and $\frac{\omega_3}{2\pi}$ of the resonant structure which are shown in the figures 2F and 2G.

Case 3: $(2 + 3)$ mode parametric resonance – It is also possible to observe the co-existence of 2 mode and 3 mode parametric resonance within a single device. Such a phenomenon is feasible if we can also locate another mode 4 which satisfies the frequency matching condition $\frac{\Omega}{2\pi} \cong \frac{\omega_4}{\pi}$. Figure 3A shows the experimental evidence for this. Here, the tones of frequencies $1.905\ MHz \cong \frac{\omega_2}{2\pi}$, $1.955\ MHz \cong \frac{\omega_3}{2\pi}$ and $1.93\ MHz \cong \frac{\omega_4}{2\pi}$ are parametrically excited (Figure 3A) when the mode of frequency $\frac{\omega_1}{2\pi}$ is driven at $\left(\frac{\Omega}{2\pi} = 3.86\ MHz\right) \cong \frac{\omega_1}{2\pi} \cong \frac{(\omega_2+\omega_3)}{2\pi} \cong \frac{\omega_4}{\pi}$ and at a higher drive level of $6\ dBm$. The vibration mode shapes of the modes 2, 3 and 4 are shown in the figures 3B-2D.

Case 4: $(3 + 3)$ mode parametric resonance – Similar to the 2 mode parametric resonance, the coexistence of multiple 3 mode parametric resonances is also possible assuming suitable vibration modes exist in such structures (Figure 3E). In a specific micromechanical device (Figure 1B), we present the evidence for this as well. Here, tones of frequencies $5.972\ MHz \cong \frac{\omega_2}{2\pi}$, $6.372\ MHz \cong \frac{\omega_3}{2\pi}$, $4.422\ MHz \cong \frac{\omega_4}{2\pi}$ and $7.922\ MHz \cong \frac{\omega_5}{2\pi}$ are parametrically excited (Figure 3E) when the mode of frequency $\frac{\omega_1}{2\pi}$ is driven at $\left(\frac{\Omega}{2\pi} = 12.344\ MHz\right) \cong \frac{\omega_1}{2\pi} \cong \frac{(\omega_2+\omega_3)}{2\pi} \cong \frac{(\omega_4+\omega_5)}{2\pi}$ and at a higher drive level of $23\ dBm$. Figures 3F-3I show the mode shapes corresponding to such parametrically excited modes 2, 3, 4 and 5. The modes 1, 2 and 3 form one 3-mode parametric resonance and the modes 1, 4 and 5 form another 3-mode parametric resonance. The observations presented in the figure 3E indicate the co-existence of these two 3-mode parametric resonances at the drive condition: $\frac{\Omega}{2\pi} = 12.344\ MHz$ (Drive frequency) and $23\ dBm$ (Drive level).

In addition to the excitation of tones associated with the modes undergoing parametric resonance, the 'Four Wave Mixing' of such tones can also be prominent owing to the elevated amplitude of such excitations. This is not possible with the case of 2 mode parametric resonance as only one tone corresponding to the sub-harmonic frequency is excited. In the 3 mode parametric resonance, the lower frequency tones $\frac{\omega_2}{2\pi}$ and $\frac{\omega_3}{2\pi}$ couples to produce $\frac{|\omega_2-\omega_3|}{2\pi}$ through the coefficient $\mu_{23}$. These in turn independently couples with $\frac{\omega_2}{2\pi}$ and $\omega_3$ through the coefficients $\mu_{12}$ and $\mu_{13}$ and the tones $\frac{\omega_2 \pm n(\omega_2 - \omega_3)}{2\pi}; n \in \mathbb{Z}$ are generated (Figure 4A). In $(2+3)$ mode parametric resonance, the subharmonic tone $\omega_4$ produced via 2 mode parametric resonance couples with the tones $\omega_2$ and $\omega_3$ produced via 3 mode parametric resonance via the additional coupling coefficients $\mu_{24}$ & $\mu_{34}$. These result in additional tones at $\frac{\omega_2 \pm n(\omega_2 - \omega_4)}{2\pi}; n \in \mathbb{Z}$ (Figure 4B). In $(3+3)$ mode parametric resonance, the tones $\omega_2$ and $\omega_3$ produced via one 3 mode parametric resonance and the tones $\omega_4$ and $\omega_5$ produced via another 3 mode parametric resonance couple through the coefficients $\mu_{23}$, $\mu_{45}$, $\mu_{24}$, $\mu_{34}$, $\mu_{25}$ & $\mu_{35}$ to generate tones at $\frac{\omega_2 \pm n(\omega_2 - \omega_3)}{2\pi}; n \in \mathbb{Z}$, $\frac{\omega_4 \pm n(\omega_4 - \omega_5)}{2\pi}; n \in \mathbb{Z}$, $\frac{\omega_4 \pm n(\omega_2 - \omega_3)}{2\pi}; n \in \mathbb{Z}$, $\frac{\omega_5 \pm n(\omega_2 - \omega_3)}{2\pi}; n \in \mathbb{Z}$, $\frac{\omega_2 \pm n(\omega_4 - \omega_5)}{2\pi}; n \in \mathbb{Z}$ and $\frac{\omega_3 \pm n(\omega_4 - \omega_5)}{2\pi}; n \in \mathbb{Z}$ (Figure 4C).

We have thus discussed the observations of $(2+2)$, 3, $(2+3)$, $(3+3)$ mode parametric resonances and the corresponding four-wave mixing processes. We now turn to the drive power level thresholds concomitant with such parametric resonances. In the singular parametric resonance cases, there exists only one drive power level threshold corresponding to its excitation. However, in the hybrid cases of parametric resonance i.e. $(2+3)$ or $(3+3)$, there may exist different drive power level threshold corresponding to each component parametric resonance. Figure 5B proves the existence of such a physical picture. When the drive power level is increased to $2\ dBm$, 2 mode parametric excitation takes place. However, with further increase to $3\ dBm$, 3 mode parametric resonance also emerges and it co-exists with the previously encountered 2 mode case. In contrast, there is only one drive power level threshold corresponding to the pure 3 mode parametric resonance (Figure 5A).

The multitude of new experimental observations falling under the umbrella of $N$ mode parametric resonance provides evidence for this paradigm. In the condensed dynamics presented in the eq. (1), there exists several system parameters including the damping coefficient, resonant frequency and coupling coefficients that control the nature of $N$ mode parametric resonance. Such dynamics can potentially trigger several new questions concerning various stochastic and deterministic traits of $N$ mode parametric resonance. For obtaining the empirical answers to such questions, it is possible to employ microelectromechanical resonators as an accessible experimental testbed.

In addition to the significance to the fundamental physics, the concept of $N$ mode parametric resonance also offers a new engineering paradigm for the field of micro and nanoelectromechanical resonators and other physical systems. Utilizing only a single weak excitation of a mechanical mode, it is possibly to parametrically excite multiple mechanical modes in this paradigm allowing access to modes that may be inefficiently coupled to in more conventional transductions schemes. Further, the onset of these different cases occur under very specific driving conditions for a given device thereby potentially enabling the engineering of functional devices utilizing this principle e.g. in signal processing and wireless encryption, multimodal acoustic sensors [24-25], mechanical computers [26-27] and also bottom-up materials engineering based tools [28-29]. Additionally, multi-mode parametric resonance can address the engineering of broadband phononic frequency combs [30-31].

Through a number of experiments on an accessible experimental testbed, the existence of multi-mode parametric resonance has been demonstrated. These results point towards the ultimate existence of $N$ mode parametric resonance in such and other similar physical systems. Such possible experimental realizations on these systems may widen the scope of $N$ mode parametric resonance and motivate the development of rigorous analytics through both phenomenology and first-principles for the prediction and estimation of $N$ mode parametric resonances in various physical systems.

## Acknowledgements
Funding from the Cambridge Trusts and UK Engineering and Physical Sciences Research Council is gratefully acknowledged.

## Authors' contributions
AG and AAS conceived the idea; AG and CD designed the device and performed the experiments; AG analysed the results; AG and AAS wrote the manuscript; AAS supervised the research.

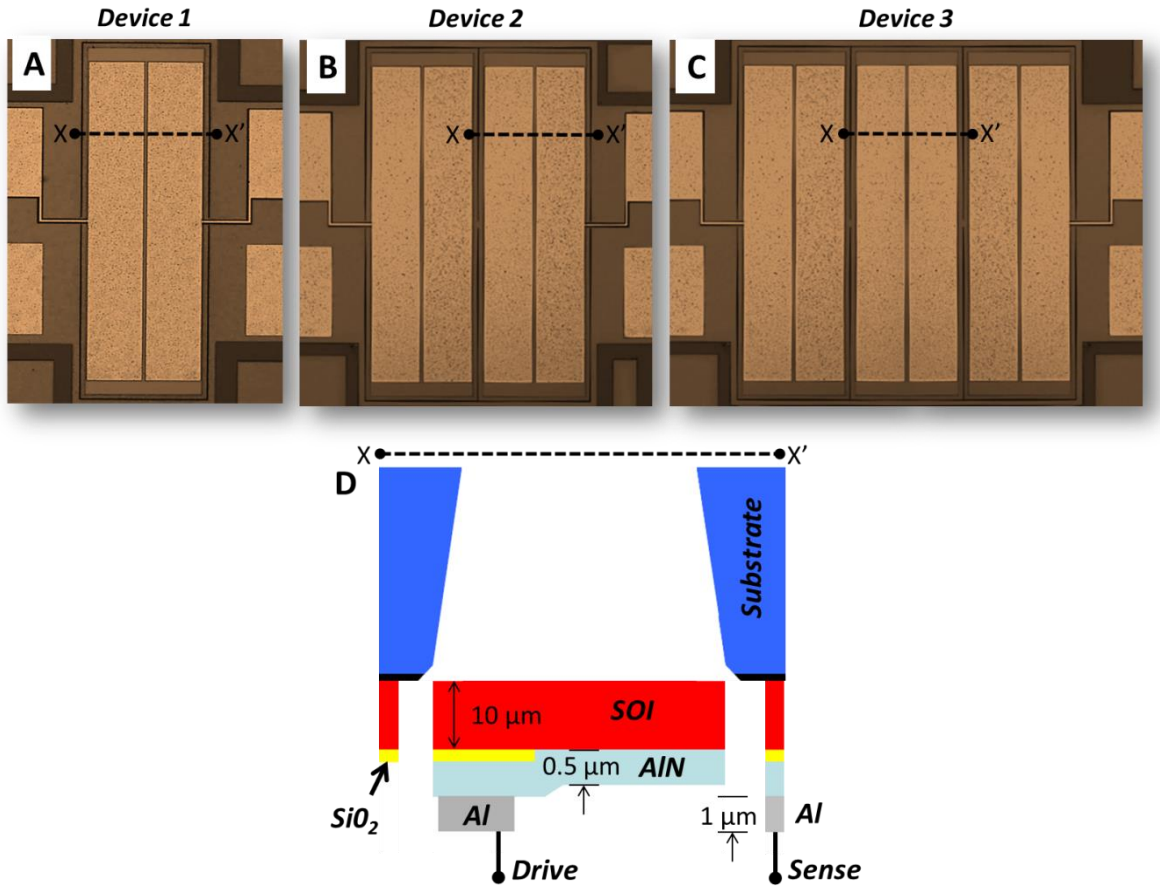

Figure 1: **Piezoelectrically driven micromechanical resonators**: **A**: Device 1: Free-free beam microstructure of dimensions $1100 \times 350 \times 11\ \mu m^3$. This has been used to demonstrate 2 and 3 mode parametric resonances; **B**: Device 2: Two free-free beam microstructures of dimensions $1100 \times 350 \times 11\ \mu m^3$ coupled by a mechanical coupler of dimensions $20 \times 2 \times 11\ \mu m^3$. This has been used to demonstrate $(3+3)$ parametric resonance; **C**: Device 3: Three free-free beam microstructures of dimensions $1100 \times 350 \times 11\ \mu m^3$ coupled by mechanical couplers of dimensions $20 \times 2 \times 11\ \mu m^3$. This has been used to demonstrate $(2+3)$ parametric resonance; **D**: Cross-sectional view of the devices A, B and C showing the layers of PiezoMUMPs fabrication process: 1 µm thick Al electrodes patterned on 0.5 µm thick AlN piezoelectric film which is in-turn patterned on SOI substrate; the 10 µm thick SOI layer is then released through back-side etch to realize mechanical functionality.

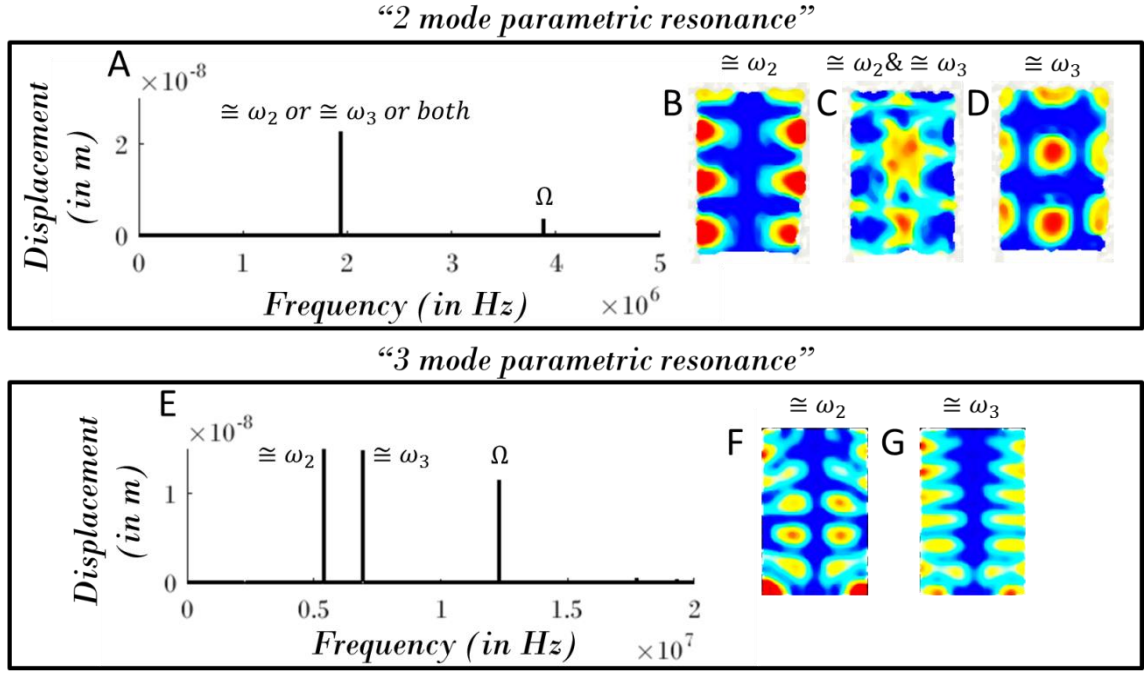

Figure 2: **$N$ Mode Parametric Resonance. A-D:** 2 Mode Parametric Resonance: **A:** Parametric excitation of tone with frequency $\frac{1}{2\pi}\frac{\Omega}{2}$; The 2-D displacement profiles at $\frac{1}{2\pi}\frac{\Omega}{2}$ when **B:** $\left(\frac{1}{2\pi}\frac{\Omega}{2} = \frac{3.862}{2} MHz\right) \cong \omega_2$; **C:** $\left(\frac{1}{2\pi}\frac{\Omega}{2} = \frac{3.874}{2} MHz\right) \cong \omega_2 \& \omega_3$; **D:** $\left(\frac{\Omega}{2} = \frac{3.876}{2} MHz\right) \cong \omega_3$. **E-G:** 3 Mode Parametric Resonance: **E:** Parametric excitation of tones with frequencies $\cong \frac{\omega_2}{2\pi}$ and $\cong \frac{\omega_3}{2\pi}$ so that their sum is equal to $\frac{\Omega}{2\pi}$; The 2-D displacement profiles at **F:** $\cong \frac{\omega_2}{2\pi}$; **G:** $\cong \frac{\omega_3}{2\pi}$. Note: The displacement profiles are normalized to the maximum displacement in the structure- i.e. Red corresponds to 1 and blue corresponds to 0.

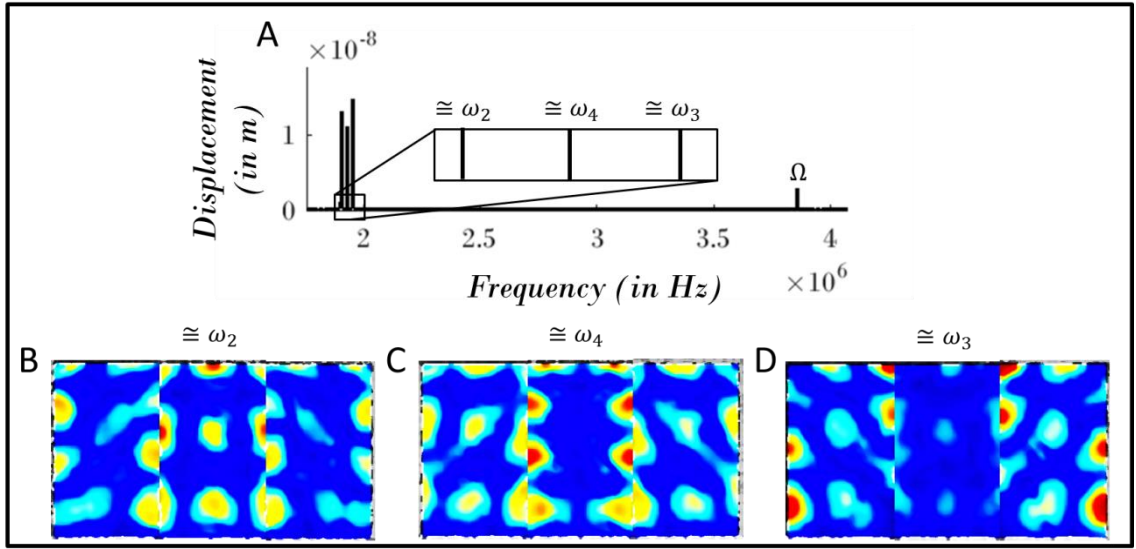

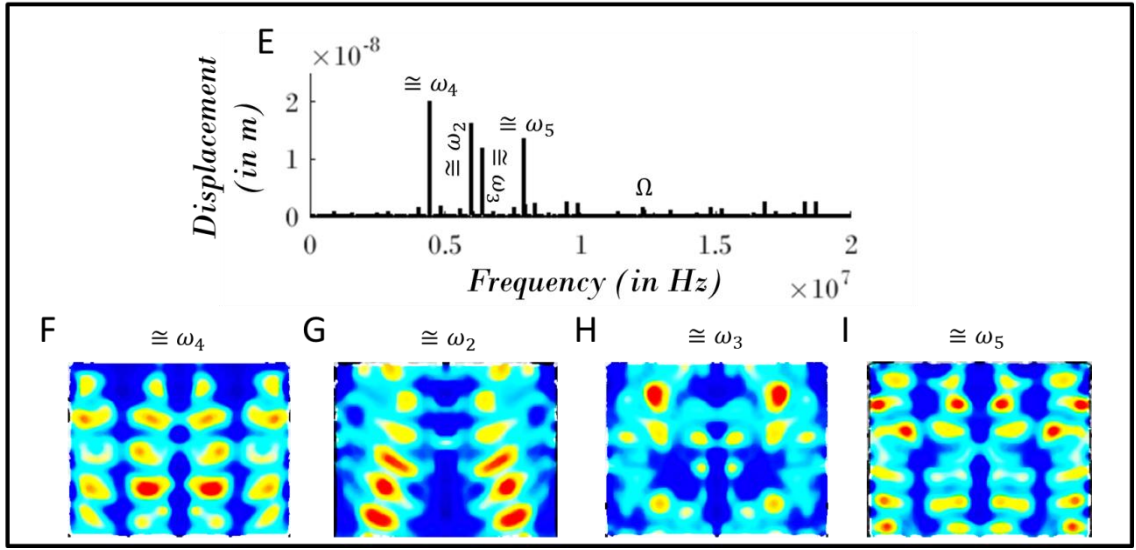

Figure 3: **Simultaneous $N$ Mode Parametric Resonances. A-D:** $(2 + 3)$ Mode Parametric Resonance: **A:** Parametric excitation of tones with frequencies $\cong \frac{\omega_2}{2\pi}$ and $\cong \frac{\omega_3}{2\pi}$ so that their sum is equal to $\frac{\Omega}{2\pi}$ and a tone with frequency $\cong \frac{\omega_4}{2\pi} \cong \frac{\Omega}{4\pi}$; The 2-D displacement profiles at **B:** $\cong \frac{\omega_2}{2\pi}$ ; **C:** $\cong \frac{\omega_4}{2\pi}$; **D:** $\cong \frac{\omega_3}{2\pi}$. **E-I:** $(3 + 3)$ Mode Parametric Resonance: **E:** Parametric excitation of tones with frequencies $\cong \frac{\omega_2}{2\pi}$ and $\cong \frac{\omega_3}{2\pi}$ so that their sum is equal to $\frac{\Omega}{2\pi}$ and tones with frequencies $\cong \frac{\omega_4}{2\pi}$ and $\cong \frac{\omega_5}{2\pi}$ so that their sum is equal to $\frac{\Omega}{2\pi}$; The 2-D displacement profiles at **F:** $\cong \frac{\omega_4}{2\pi}$ ; **G:** $\cong \frac{\omega_2}{2\pi}$; **H:** $\cong \frac{\omega_3}{2\pi}$; **I:** $\cong \frac{\omega_5}{2\pi}$. Note: The displacement profiles are normalized to the maximum displacement in the structure- i.e. Red corresponds to 1 and blue corresponds to 0.

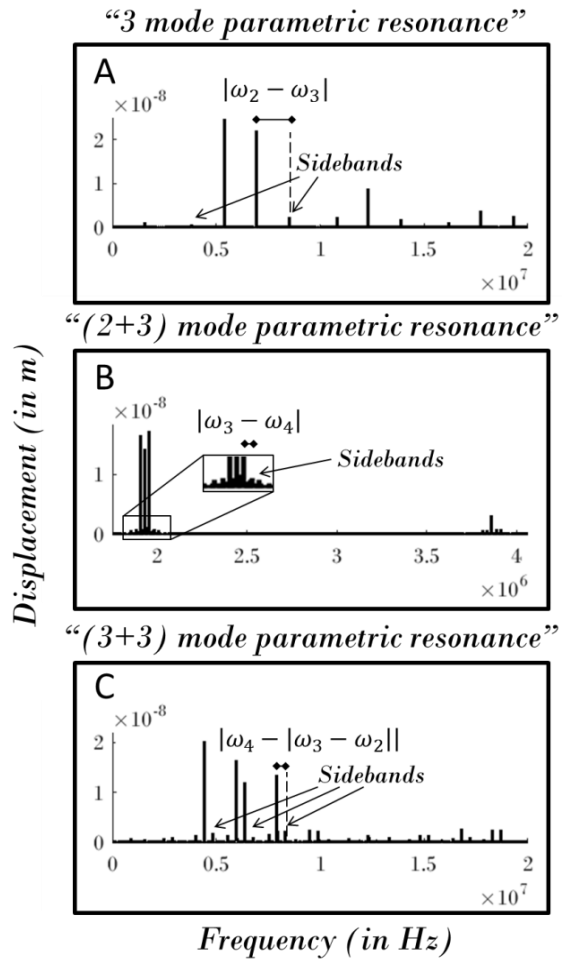

Figure 4: **Four-Wave Mixing of Parametric Excitations. A:** The sidebands with spacing $|\omega_2 - \omega_3|$ are generated in 3 mode parametric resonance. The drive frequency $\frac{\Omega}{2\pi} = 12.331\ MHz$ and the drive level $P = 18\ dBm$; **B:** The sidebands with spacing $|\omega_3 - \omega_4|$ are generated in the $(2 + 3)$ mode parametric resonance. The drive frequency $\frac{\Omega}{2\pi} = 3.86\ MHz$ and the drive level $P = 8\ dBm$; **C:** The sidebands with spacing $|\omega_4 - |\omega_3 - \omega_2||$ are generated in $(3 + 3)$ mode parametric resonance. The drive frequency $\frac{\Omega}{2\pi} = 12.344\ MHz$ and the drive level $P = 23\ dBm$.

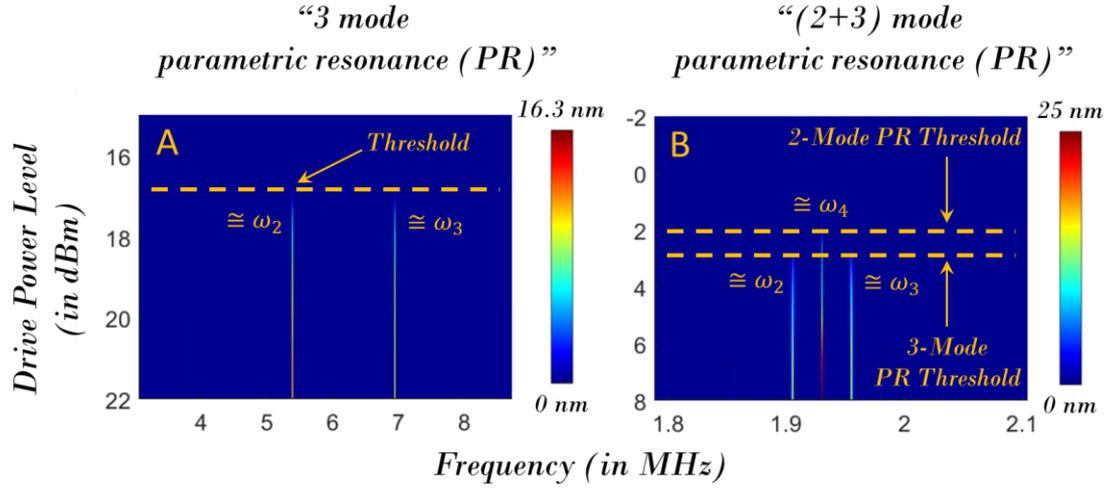

Figure 5: **Drive Power Level Dependence of Parametric Resonances.** Frequency contours for different drive power levels. Here, the Red and blue correspond to the maximal and minimal displacement amplitudes respectively. **A:** 3 Mode Parametric Resonance: Parametric excitation of tones with frequencies $\cong \frac{\omega_2}{2\pi}$ and $\cong \frac{\omega_3}{2\pi}$ so that their sum is equal to the drive frequency $\frac{\Omega}{2\pi}$; **B:** (2 + 3) Mode Parametric Resonance: Parametric excitation of tones with frequencies $\cong \frac{\omega_2}{2\pi}$ and $\cong \frac{\omega_3}{2\pi}$ so that their sum is equal to the drive frequency $\frac{\Omega}{2\pi}$ and a tone with frequency $\cong \frac{\omega_4}{2\pi} \cong \frac{\Omega}{4\pi}$.

**Supplementary Information**

**Towards *N* mode parametric electromechanical resonances**

Authors: Adarsh Ganesan[1], Cuong Do[1], Ashwin Seshia[1]

[1] Nanoscience Centre, University of Cambridge, Cambridge, UK, CB3 0FF

**Supplementary section S1**

Reduced dynamics for each case of parametric resonance

$(2+2)$ mode parametric resonance

$$\ddot{Q}_1 + \omega_1^2 Q_1 + 2\zeta_1\omega_1\dot{Q}_1 + \sum_{i=2}^{N}\mu_{ii}Q_i^2 + P_1 \cos\Omega t = 0$$
$$\ddot{Q}_2 + \omega_2^2 Q_2 + 2\zeta_2\omega_2\dot{Q}_2 + \mu_{12}Q_1 Q_2 = 0 \quad \text{(S1)}$$
$$\ddot{Q}_3 + \omega_3^2 Q_3 + 2\zeta_3\omega_3\dot{Q}_3 + \mu_{13}Q_1 Q_3 = 0$$

3 mode parametric resonance

$$\ddot{Q}_1 + \omega_1^2 Q_1 + 2\zeta_1\omega_1\dot{Q}_1 + \mu_{23}Q_2 Q_3 + P_1 \cos\Omega t = 0$$
$$\ddot{Q}_2 + \omega_2^2 Q_2 + 2\zeta_2\omega_2\dot{Q}_2 + \mu_{13}Q_1 Q_3 = 0 \quad \text{(S2)}$$
$$\ddot{Q}_3 + \omega_3^2 Q_3 + 2\zeta_3\omega_3\dot{Q}_3 + \mu_{12}Q_1 Q_2 = 0$$

$(2+3)$ mode parametric resonance

$$\ddot{Q}_1 + \omega_1^2 Q_1 + 2\zeta_1\omega_1\dot{Q}_1 + \mu_{23}Q_2 Q_3 + \mu_{44}Q_4^2 + P_1 \cos\Omega t = 0$$
$$\ddot{Q}_2 + \omega_2^2 Q_2 + 2\zeta_2\omega_2\dot{Q}_2 + \mu_{13}Q_1 Q_3 + \mu_{24}Q_2 Q_4 = 0$$
$$\ddot{Q}_3 + \omega_3^2 Q_3 + 2\zeta_3\omega_3\dot{Q}_3 + \mu_{12}Q_1 Q_2 + \mu_{34}Q_3 Q_4 = 0 \quad \text{(S3)}$$
$$\ddot{Q}_4 + \omega_4^2 Q_4 + 2\zeta_4\omega_4\dot{Q}_4 + \mu_{14}Q_1 Q_4 = 0$$

$(3+3)$ mode parametric resonance

$$\ddot{Q}_1 + \omega_1^2 Q_1 + 2\zeta_1\omega_1\dot{Q}_1 + \mu_{23}Q_2 Q_3 + \mu_{45}Q_4 Q_5 + P_1 \cos\Omega t = 0$$
$$\ddot{Q}_2 + \omega_2^2 Q_2 + 2\zeta_2\omega_2\dot{Q}_2 + \mu_{13}Q_1 Q_3 = 0$$
$$\ddot{Q}_3 + \omega_3^2 Q_3 + 2\zeta_3\omega_3\dot{Q}_3 + \mu_{12}Q_1 Q_2 = 0 \quad \text{(S4)}$$
$$\ddot{Q}_4 + \omega_4^2 Q_4 + 2\zeta_4\omega_4\dot{Q}_4 + \mu_{15}Q_1 Q_5 + \mu_{24}Q_2 Q_4 + \mu_{34}Q_3 Q_4 = 0$$
$$\ddot{Q}_5 + \omega_5^2 Q_5 + 2\zeta_5\omega_5\dot{Q}_5 + \mu_{14}Q_1 Q_4 + \mu_{25}Q_2 Q_5 + \mu_{35}Q_3 Q_5 = 0$$

Here, $\frac{\omega_i}{2\pi}$ and $\zeta_i$ are the natural frequency and damping coefficient respectively, $\mu_{nm}$ are the quadratic coupling coefficients and $P_1 \cos \Omega t$ is the drive.